\documentstyle[prl,aps,epsf]{revtex}
\begin{document}
\draft
\def\bfs{{\bf s}}   \def\bfp{{\bf p}}
\def\calS{{$\cal S$}} \def\calP{{$\cal P$}} \def\calSp{{$\cal S'$}}
\def\calL{{$\cal L$}} 
\def\nmax{{n_{max}}} \def\nsb{{n_{10}}}
\def\dh{{$d(\bfs_1, \bfs_2)$}}
\def\dhm{{$d_{min}(\bfs_1, \bfs_2)$}}
\def\dn{{$\Delta n_{10}$}}
\def\blackbox{{\rule{2.mm}{2.mm}}}
\def\Bull{{\Large\vrule height .9ex width .8ex depth -.1ex}}
\twocolumn[\hsize\textwidth\columnwidth\hsize\csname %
@twocolumnfalse\endcsname

\title{Geometric and Statistical Properties of the Mean-Field HP
Model, the LS Model and Real Protein Sequences}
\author{C.T. Shih$^{1,6}$, Z.Y. Su$^1$, J.F. Gwan$^{2}$,
B.L. Hao$^{3}$, C.H. Hsieh$^1$, J.L. Lo$^4$ and H.C. Lee$^{4,5}$}
\address{
$^{1}$National Center for High-Performance Computing, Hsinchu, Taiwan, ROC\\
$^{2}$Forum Modellierung, Forschungszentrum J$\ddot{u}$lich, D 52425 
J$\ddot{u}$lich, Germany\\
$^{3}$ Institute of Theoretical Physics, Academia Sinica, Beijing, China\\
$^{4}$Dept. of Physics and Dept. of Life Science, 
National Central University, Chungli, Taiwan, ROC\\
$^{5}$Department of Physics, Stanford University, Palo Alto, CA 94305, USA\\
$^{6}$Department of Physics, Tunghai University, Taichung, Taiwan, ROC}

\date{\today}
\maketitle
\begin{abstract}
Lattice models, for their coarse-grained nature, are best suited 
for the study of the ``designability problem'', the phenomenon in which  
most of the about 16,000 proteins of known structure have their 
native conformations concentrated in a relatively small number 
of about 500 topological classes of conformations.  Here it is 
shown that on a lattice the most highly designable simulated 
protein structures are those that have the largest number of 
surface-core switchbacks.  A combination of physical, mathematical 
and biological reasons that causes the phenomenon is given.  By 
comparing the most foldable model peptides with protein sequences in 
the Protein Data Bank, it is shown that whereas different models may 
yield similar designabilities, predicted foldable peptides will simulate  
natural proteins only when the model incorporates the correct physics 
and biology, in this case if the main folding force arises from the 
differing hydrophobicity of the residues, but does not originate, say, 
from the steric hindrance effect caused by the differing sizes of the 
residues.  
\end{abstract}
]

\section{Introduction}
\label{s:intro}

It is believed that the dynamical
folding of a protein to its native conformation
is determined by the 
amino acid sequence of the protein \cite{anfinsen73}. Yet 
the folding of any particular protein is an extremely complex
process; simulation of the folding of even a small protein 
remains an unsurmounted challenge to state-of-the-art computers \cite{duan}. 
Nevertheless, a good
understanding of a number of general features of protein folding have
been acquired in computational studies using simple lattice models
\cite{dill,shak94,funnel,li96,li98,shak98}.  One feature is the so-called
funnel picture that leads to a two-state description of folding
\cite{funnel,visual_funnel}.  Here the vertical dimension of the
funnel represents the state of foldedness of the protein (or roughly
its free energy), which increases (decreases) from the top towards the
bottom of the funnel, and a cross-section of the funnel represents the
conformation space accessible to the folding protein at a given state
of foldedness.  Near the top of the funnel, most conformations are
freely accessible and folding proceeds extremely rapidly.  As the
folding progresses and the opening of the funnel narrows, accessibility
of one conformation from another becomes increasing restrictive, so
that increasingly fewer pairs of conformations are connected by
almost-equal-energy paths and folding correspondingly slows down.  An
alternative view is that the energy landscape becomes increasingly
rugged.  At some junction the rate of decrease in the number of
accessible conformations, hence the rate of decrease in entropy, is
so large as to cause the rate of free-energy change as a function of
foldedness to be positive, so that a free-energy barrier is formed
to become an obstacle against further folding.  
At this point folding practically grinds to
halt and can proceed stochastically only on very rare occasions 
that brings it over the barrier, after which the protein folds (and
unfolds) relatively rapidly 
to its native conformation in an annealing-like process.
 
Another issue clarified by simple lattice models is the designability
of "topological" classes of protein conformations
\cite{li96,li98,shih00}.  The designability of a conformation class
is the number of proteins whose native conformations belong to the
class.  At the moment the number of proteins with known
three-dimensional conformations in the Protein Databank (PDB
\cite{PDB}) is of the order of 16,000 and is increasing rapidly, while
the number of conformation classes has remained about 500 for some time
and is not expected to grow beyond 1000.  Even when the the fact that
many proteins in the PDB are homologues with similar structures are
taken into account, the discrepancy between the number of
non-homologous proteins and the number of conformation classes 
of observed native conformations is glaring.  
Because a class is in fact composed of many conformations that
differ in detail (such differences could very well be
important to the function of proteins), 
the problem of designability is best studied in
coarse-grain models, such as lattice models, that disregard such
details.

The simplest interacting lattice model is the 
HP model proposed by Dill {\it et al.} \cite{dill}, 
in which the 20 kinds of amino acids are divided into two types, 
hydrophobic (H) and polar (P). 
This model has been studied extensively by several groups
in the last decade \cite{dill,shak94,funnel,li96,li98,shak98}. 
A mean-field version of the model that yields tremendous 
simplification was used to study the designability problem, and 
it was found that the designabilities of structures vary 
greatly (the terms structures and conformation classes will be used 
interchangeably in this paper), and that only a 
tiny portion of structures are highly designable.  Moreover, 
it was noted that highly designable structures seem to have patterns 
that emulates secondary structural motifs \cite{li96,li98,shih00}.

In a general Hamiltonian setting, the Hamiltonian $\cal H$ can be
viewed as a mapping of the peptide space $\cal P$ to the conformation
space $\cal C$.  When $\cal C$ is sufficiently coarse grained, which
is the case we consider, each point in $\cal C$ is a topological class
of native conformations.  Then $\cal H$ is a mapping of $\cal P$ to
such conformation classes into $\cal C$.  If we remove from $\cal P$ all
the peptides that are mapped by $\cal H$ to more than one conformation
class in $\cal C$ (i.e., the degenerate cases), 
the remainder of $\cal P$ is partitioned by $\cal
H$ into equivalent classes of peptides, with each peptide class being
mapped to a single conformation class.  Designability
results from a highly skewed distribution of the {\it size} of the
peptide classes.   We shall call peptides belonging to 
peptide classes that are mapped to highly designable structures 
highly foldable peptides.  

In \cite{li98} the designability issue of the mean-field HP model 
was reduced to a purely geometric problem which rendered it 
easy to discuss and visualize the skewed distribution 
of the size of peptide classes.   It was however 
not made clear what characterizes those structures that are 
highly designable, nor was it demonstrated whether or not 
highly foldable peptides have anything to do with real proteins.  
In fact, whereas one can well imagine many 
$\cal H$'s in lattice models to yield biased designability, 
it is not clear that any such $\cal H$ would yield foldable 
peptides that simulate real proteins. 

In this paper, expanding on claims made in an earlier letter
\cite{shih00}, the highly designable structures in the mean-field HP
model will be characterized - they are those that have the largest
number of surface-core switchbacks, and it will be shown that 
highly foldable peptides have a high similarity with
real protein sequences in general and with segments of sequences that
fold to $\alpha$ helices in particular. 

To demonstrate a point made above, this paper also discusses a lattice
model that exhibits designability but does not seem to be biologically
correct.  In the LS model, the 20 kinds of amino acids are divided
into two types, large (L) and small (S), and it is assumed that the
deciding factor in folding is the the steric hindrance effect caused
by the difference in the sizes of the amino acids \cite{micheletti98}.
It was shown in ref. \cite{micheletti98} that on a lattice, structures
in the LS model too have uneven designability (there called
encodability score); only a small portion of structures, also claimed
to have protein-like secondary structures, are selected by large
numbers of peptide sequences as unique ground states.  It will be
shown here that in spite of the fact that the LS model is
mathematically almost equivalent to the mean-field HP model, unlike
the mean-field HP model, highly foldable peptides 
in the LS model do not match well with real protein sequences.

In the following two sections the mean-field HP model and the LS model
are reviewed and it is shown that, notwithstanding their quite 
different physical contents, on square lattices the two 
models are mathematically close  approximates.  In Section 4 
the geometrical properties of a two-dimensional square lattice 
and the way they restrict the space of structures, 
which are compact paths on the lattices, are discussed.  In Section 5 
it is shown that only a very small portion of the structure have the 
highest numbers of surface-core switchbacks and that, for both models, 
it is these structures that have the highest designabilities.  
Because the partition of amino acids in the HP model is based on 
hydrophobicity while that in the LS model is based on residue size, 
the highly foldable peptides are translated into different sets 
of ``physical'' peptides in the two models.  
In Section 6 the highly foldable peptides in the two models are 
compared with real proteins in the Protein Data Bank and it is 
shown that the highly foldable peptides in the HP model match well 
with real protein sequences in general and with segments of sequences 
that fold to $\alpha$ helices in particular (but not well 
with segments of sequences that fold to $\beta$ sheets), whereas 
those in the LS model match poorly with real protein sequences. 
Section 7 gives an expanded discussion of our results. 
In an Appendix the most highly foldable peptides in the two models 
are given and compared. 

\section{The HP Model}
\label{s:model}

The Hamiltonian of the HP model is:
\begin{equation}
H=\sum_{i<j} E_{p_i p_j} \Delta(\vec{r_i}-\vec{r_j})
\label{e:hp1}
\end{equation}
where $p_i$ is the type, H for hydrophobic and P for polar, 
of the $i$th residue, or amino acid, 
in the peptide chain \cite{dill}; $\Delta(\vec{r_i}-\vec{r_j})=1$ if
$\vec{r_i}$ and $\vec{r_j}$ are nearest neighbors in the
lattice but not adjacent along the peptide sequence, and
$\Delta(\vec{r_i}-\vec{r_j})=0$ otherwise; $E_{p_i p_j}$
specifies the residue contact energies that depend on the
types of residues in contact.

Several sets of contact energies $(E_{HH},E_{HP},E_{PP})$ have
been used: $(-1,0,0)$ for the original HP model \cite{dill},
$(-2.3,-1,0)$ by Li {\it et al.} \cite{li96}, and $(-\pi,-1,0)$
by Buchler and Goldstein \cite{buchler99}.  Li {\it et al.}
suggested that the contact energies should satisfy the following
constraints: 1) compact shapes have lower energies than 
non-compact shapes; 2) $E_{PP}>E_{HP}>E_{HH}$ so that hydrophobic
residues are buried as much as possible; and 3) different types
of residues tend to segregate, which is a condition induced by 
having $2E_{HP}>E_{PP}+E_{HH}$ \cite{li96,li97}.  In this work these 
will be adopted with the modification that 3) is replaced by the 
additive relation $2E_{HP}=E_{PP}+E_{HH}$.  
Then the potential simplifies to:
\begin{equation}
E_{p_i p_j}=-(p_i +  p_j)
\label{e:additive}
\end{equation}
where $p_i=1$ for H and $p_i=0$ for P residue \cite{ejtehadi98}. 
Henceforth only structures that correspond to self-avoiding 
compact paths on a lattice will be considered. 

In an $N$$\times$$N$ two-dimensional square lattices, there are
four corner sites with coordination number $N_n=2$, $4(N-2)$
side sites with $N_n=3$ and $(N-2)^2$ core sites with $N_n=4$. 
With the exception of the two ends of the peptide chain, which 
we ignore, each lattice point has $N_n-2$ contacts. 
So the Hamiltonian Eq.(\ref{e:hp1})
becomes:
\begin{eqnarray}
H &=& -(0\times\sum_{i\in corner} + 1\times\sum_{i\in side} +
   2\times\sum_{i\in core})p_i\nonumber\\
  &=& -\sum_i p_i - \sum_{i\in core} p_i +
   \sum_{i\in corner} p_i
\label{e:hp1.5}
\end{eqnarray}
The first term on the right-hand side of Eq.(\ref{e:hp1.5}) is a
constant for a given peptide sequence.  It is independent of whatever 
conformation the peptide resides in and, since Eq.(\ref{e:hp1.5}) 
will only be used here to  determine the native structure 
of a particular peptide sequence, it will be omitted.   The third term
means that it is costly to put H residues in the corner sites. 
Since it is of order $1/N^2$ it too will be omitted.  
The Hamiltonian then simplifies to what is known as the mean-field HP 
model \cite{li98}:
\begin{equation}
H({\bf p}, {\bf s})  =  -{\bf p}\cdot{\bf s} =
\frac12 (|{\bf s} -{\bf p}|^2 - {\bf p}^2 - {\bf s}^2)
\label{e:hp2}
\end{equation}
where ${\bf p}= (p_1,p_2,\ldots,p_n)$, 
$n=N^2$, is the binary peptide sequence
and ${\bf s}= (s_1,s_2,\ldots,s_n)$ is a binary structural sequence
converted from a self-avoiding compact path on the lattice 
with the assignment: $s_i= 1$ (0)
if the $i$th site of the structure is a core (surface) site. 
In this new form the Hamiltonian has an interpretation quite different 
from its original meaning.  There it was an expression of inter-residual 
interaction.  Here in Eq.~(\ref{e:hp2}) it is no longer inter-residual, 
rather it has the form of a site-dependent potential.   
With ${\bf s}^2$ fixed for a given lattice and ${\bf p}^2$ a constant 
for a given peptide sequence, both are irrelevant to
the determination of the ground state structure of the peptide. 
They will be ignored in the ensuing calculation. 
The Hamiltonian now reduces to one-half of $|{\bf s} -{\bf p}|^2$ 
and a neat geometric interpretation for it emerges  \cite{li98}.  
When  ${\bf p}$ and ${\bf s}$ are viewed as $n$-component vectors, 
this quantity is just the Hamming distance between two corner 
points in a unit $n$-dimensional hypercube. 

When the energy matrix elements are not additive,
that is, when $E_{HH}=-2-\gamma$ with $\gamma>0$ as was used in 
\cite{dill,li96,buchler99},  the model cannot be reduced to the simple
site-dependent form of Eq.(\ref{e:hp2}).
The effect of $\gamma$ is to stabilize the low-lying states in 
the mean-field model further by increasing the number of H-H contacts.

\section{The LS Model}
\label{s:LSmodel}

It was shown by Micheletti {\it et al.} that in the LS model 
the designability (called encodability score by the authors) 
distribution of structures is similar to that in the mean-field 
HP model \cite{micheletti98}.
The Hamiltonian of this model is
\begin{equation}
H=-\sum_i z_i(\Gamma)\cdot A(z(\sigma_i)-z_i(\Gamma))
\label{e:lsmodel}
\end{equation}
where $\sigma_i \in \{L,S\}$; $z(\sigma_i)$ is the maximal number of 
nearest contacts without steric repulsion belonging to residue $i$; 
on a square lattice, 
$z(\sigma_i)$ is equal to 1 (2) for $L$ ($S$) residues inside the chain, 
and to 2 (3) for $L$ ($S$) residues at chain ends; 
$z_i(\Gamma)$ is the number of contacts
of the $i$th residue in a conformation $\Gamma$; and $A(x)$ equals to
1 if $x\ge 0$ and $-a<0$ otherwise.  
The Hamiltonian implies that 
if the number of contacts of the $i$th residue is larger than
$z(\sigma_i)$, then the contact energy will be increased by $a$ 
owing to steric effects. 

\begin{minipage}{3.2in}
\begin{figure}[m]
\epsfysize=3.0cm\epsfxsize=3.2in\epsfbox{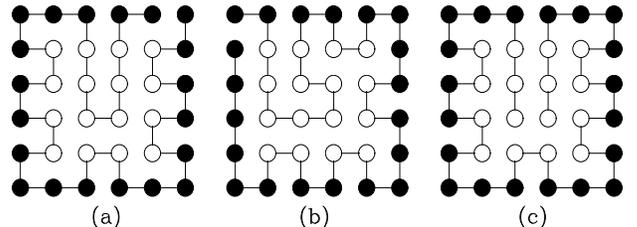}
\caption{(a) The most (third most) designable, 
(b) the second most (most) designable  and (c)
the third (second) most designable structures in 
the mean-field HP (LS) model, respectively, on a 6$\times$6 lattice.}
\label{f:struct2}
\end{figure}
\end{minipage}

\bigskip
The results in Ref. \cite{micheletti98}, where 
$a$ was set equal to $\infty$, show that the distribution
of designability of structures in LS model
is very similar to that in the HP model.   In fact most of the 
highly designable structures in one model 
are likewise in the other model (see Appendix).  
The highly designable structures 
in the LS model also have protein-like secondary substructure
and tertiary symmetries.
Three among the most designable structures in the two 
models are shown in Fig.~\ref{f:struct2}.

Just as practiced in the last section, we consider only 
compact structures and neglect the effect of the two end points 
on a peptide chain.  Table~\ref{t:LS_Hamiltonian} gives the 
values of $x$, $A(x)$ and Hamiltonian for the two types of 
residues at corner, side and core sites on a square lattice.  
Let $o$, $s$ and $c$ denote the number of corner, side and core sites, 
respectively; $n$ = $o$ + $s$ + $c$ = $N^2$ the total number of sites; 
and the subscripts $L$ and $S$ denote residue type, then 
\begin{eqnarray}
H &=& - s_L + 2a c_L -  s_S - 2 c_S \nonumber\\
  &=& 2an_L - (1+2a) s_L - 2a o_L - n_S - c_S + o_S
\label{e:ls1.5}
\end{eqnarray}

For a given peptide sequence, $n_L$ and $n_S$ are fixed.  
First consider the case when the steric repulsion is strong 
but finite, namely, $a\gg 1$.  Dropping the corner term $o_S$ one gets
for a given peptide sequence, 
\begin{equation}
H = -(2a+1)  c_S + {\rm const.} 
  \approx -2a{\bf p}\cdot{\bf s} + {\rm const.}
\label{e:ls2}
\end{equation}
where {\bf p} and {\bf s} are the peptide and structure binary 
vectors defined before, with the exception that in {\bf p} the digit 
0 (1) now stands for L (S).   Comparison of this equation with 
Eq.~(\ref{e:hp2}) reveals that, at least on a square lattice, the
mathematical form of the two models are essentially identical,  
provided that here the pair H and P in the HP model is replaced 
by S and L, respectively.  Since there is only one scale in 
either model, the size of $a$ does not matter so far as it is 
much greater than unity but finite.  

\begin{minipage}{3in}
\begin{table}[m]
\caption{Action of the Hamiltonian for the LS model on a square lattice;
end points of chains are ignored and $x=z(\sigma)-z(\Gamma)$.}
\begin{tabular}{ccccc}
 type&&corner&side&core \\
\hline
 &$z(\Gamma)$&0&1&2 \\
\hline
 &$x$&2&1&0 \\
  S&$A(x)$&1&1&1 \\
 &$H$&0&-1&-2 \\
\hline
 &$x$&1&0&-1 \\
 L&$A(x)$&1&1&-a \\
 &$H$&0&-1&2a \\
\end{tabular}
\label{t:LS_Hamiltonian}
\end{table}
\end{minipage}

When $a\to\infty$, as was the case in \cite{micheletti98}, the 
term $2a c_L$ in the first line of  Eq.~(\ref{e:ls1.5}) 
becomes a constraint that L residues are prohibited from 
core sites, namely $c_L=0$ strictly, and the rest of the 
Hamiltonian becomes 
\begin{equation}
H = -c_S + o_L -n_L + o_S -n_S \approx 
      - {\bf p}\cdot{\bf s} + {\rm const.}
\label{e:ls3}
\end{equation}
which again coincides with Eq.~(\ref{e:hp2}). 

\vspace{-12pt}
\section{Geometrical Properties of the 2D Square Lattice}
\label{s:geometric}

Since Eqs.~(\ref{e:hp2}), (\ref{e:ls2}) and (\ref{e:ls3}) 
reduce the Hamiltonians  
of the mean-field HP and LS models to the same problem in geometry, 
namely one of the Hamming distance between the two vectors \bfs\ and \bfp, 
we now study the space of these vectors (in the HP model).  
Consider an $N$$\times$$N$ square 
lattice with $n =N^2$ sites.  Recall that every  structure is a 
self-avoiding compact path on the lattice. 
The set \calP\ of all binary peptides \bfp\ is then 
just the set of $2^{n}$ binary sequences.  Because of geometric 
constraints, the set \calS$\subset$\calP\ 
of binary structure sequences \bfs\ is 
far smaller than \calP.    For a very rough estimate for the upper 
limit of the size of \calS, consider the construction of compact paths 
by random walk on the lattice.  At any given point during the walk 
after the first step, the maximum number of allowed next steps is the 
coordination number minus one, which is between 2 and 3.   
As the number of steps taken increases, the average number  of allowed 
next steps will decrease.   We take the average number to be 2 
up to the point when the lattice is half full.
For a randomly chosen path, 
after the lattice is half full, chances are that the number of allowed 
next steps will be either one or zero most of the time. 
So the number of allowed  \bfs' should be much less than $2^{n/2}$.  
On a 6$\times$6 lattice this last number is 262144,   
whereas the size of \calS\ is  30408, and the  
size of \calP\ is $2^{36}=68,719,476,736$. 
An example of an allowed {\bf s} on the 6$\times$6 lattice
is shown in Fig.~\ref{f:struct1} (a). 
If we think of \calP\ as the set of all the corner points 
in the $n$-dimensional unit hypercube, then the set \calS\ is composed 
of a tiny subset of corner points.  It was shown earlier that 
the designability of an {\bf s} $\in$ \calS\ is the Voronoi polytope 
of {\bf s} in \calP; it is clear what characterizes the 
designability problem is the distribution of the contents 
of \calS\ in the  unit hypercube.

We now examine how geometric constraints reduce \calP\ down 
to \calS.  A sequence in  \calP\
may be viewed as a chain of 0's and 1's connected by $n-1$ links of three
types, those connecting 0 and 0 sites, 0 and 1 or 1 and 0 sites, and
1 and 1 sites, respectively.     Let the numbers of such links be  
$n_{00}$, $n_{10}$ and $n_{11}$, respectively. 
The sequence is partitioned by
the 1-0 links into $n_{10}+1$ segments of contiguous 1's or
0's.  Whereas the link numbers for a \bfp\ are devoid of geometric
meaning, that for \bfs\ are the consequences of geometric constraints.
To illustrate this, consider the case 
$N>4$ (the surface to core ratio in smaller lattices are too lop-sided
to be of interest).  Some of the simplest constraints that must be 
satisfied by an allowed \bfs\ are:

\begin{enumerate}
\item An isolated single 0 may only occur at an end of a path;

\vspace{-5pt}\item An isolated single 1 may only either occur at or be one
0-segment away from an end of a path;

\vspace{-5pt}\item Each of the four corners on the lattice belongs 
to a 0-segment at least 4 sites long, except when the corner is an end 
of a path;

\vspace{-5pt}\item For a path having the pattern \bfs~$=(1\cdots1)$ 
(both the ends of the path are 1-sites),  
$2n_{00}+n_{10}= 8N-8$ and $2\le n_{10} \le 4N-12$;

\vspace{-5pt}\item For \bfs~$=(0010011\cdots1)$,  
$2n_{00}+n_{10}= 8N-9$ and $5\le n_{10}\le 4N-11$;

\vspace{-5pt}\item For \bfs~$=(0010011\cdots1100100)$, 
$2n_{00}+n_{10}= 8N-10$ and $10\le n_{10}\le 4N-10$ if $N>6$,   
the last relation is replaced by $8\le n_{10}\le 4N-10$ if $N\le 6$; 

\vspace{-5pt}\item For \bfs~$= (0010011\cdots 0)\ne (0010011\cdots1100100)$,  
$2n_{00}+n_{10}= 8N-10$ and $4\le n_{10}\le 4N-12$; 

\vspace{-5pt}\item For \bfs~$= (0\cdots0)\ne(0010011\cdots 0)$ and 
$\ne (0010011\cdots1100100)$,
$2n_{00}+n_{10}= 8N-10$ and $2\le n_{10}\le 4N-12$; 

\vspace{-5pt}\item For \bfs~$= (0\cdots1)\ne(0010011\cdots1)$,  
$2n_{00}+n_{10}= 8N-9$ and $1\le n_{10}\le 4N-13$. 
\end{enumerate}

\noindent The first two rules are obvious on a square lattice.
The third rule implies that the polar residues tend to accumulate
around corners. This fortuitously reflects a property of
real proteins: the relative abundance of polar residues on surface
areas with large curvatures.  Figs.~\ref{f:struct1} (b) and (c) 
illustrate the origin of the fourth rule on a 6$\times$6 lattice.   
The two structures are both of the type $(1\cdots 1)$, that is, they 
begin and end both on core sites.   The dark solid 
links in the figures define ``templates'' for constructing \bfs'  
that respectively have the 
maximum (twelve) and  minimum (two) values for $n_{10}$. 
Rules (5)-(8) can be shown in a similar way.
By explicitly applying the above rules in the selection of \bfs\ 
(as opposed to requiring an \bfs\ to be a compact self-avoiding path), 
the total number of $2^{36}=68,719,476,736$ binary sequences in \calP\
is reduced to a set of  $537549$ candidate paths which, 
relatively speaking, is now only slightly 
greater than the exact number ($30408$) of \bfs' in \calS. 
This implies that the set of rules given above embodies the essence 
of the geometric requirement that guarantees elements in  \calS\ 
to be compact self-avoiding paths. 

\begin{minipage}{3.2in}
\begin{figure}[m]
\epsfysize=3cm\epsfxsize=3.2in\epsfbox{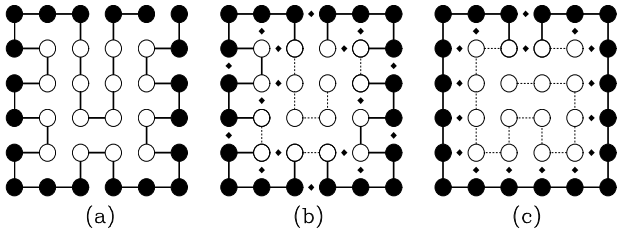}
\caption{(a) A structure defined by a compact, self-avoiding path, 
which is in turn represented by the binary sequence 
(001100 110000 110000 110011 000011 111100).  Black (white) discs
represent surface (core) sites coded by the digit 0 (1). In (b) and (c),
the dark, solid links define ``templates'' for constructing structures 
of the type $(1\cdots 1)$ whose $n_{10}$ values 
are 12 and 2, respectively.} 
\label{f:struct1}
\end{figure}
\end{minipage}

\vspace{-10pt}
\section{Distribution of the Allowed Structures in the Hypercube}
\label{s:statistics}

Here we show that only a small portion of the structures in 
\calS\ have large $n_{10}$.  
On an $N$$\times$$N$ square lattice, there is a total of $2N^2-2N$
links and $N^2-1$ among them need to be chosen to form a structure. 
For the 6$\times$6 case these numbers are 60 and 35, respectively. 
For the structure shown in Fig.~\ref{f:struct1} (b),  of the 
total number of 60 links on the lattice, 28 links 
are used to define the template (that has $n_{10}$=12) 
and 17 links, marked by filled diamonds in the figure, are forbidden 
because they would form close loops or connect sites which already
have two links.  This means that to complete an \bfs\ from 
the template, one needs to select  $35-28=7$ links from among 
$60-28-17=15$ links on the lattice.  Hence at most ${15\choose 7} = 6435$ 
\bfs' with $n_{10}$~=~12 can be constructed from the template. 
A similar argument shows that ${23\choose 14} = 817190$ 
\bfs' with $n_{10}$~=~2 can be constructed from the template shown 
in Fig.~\ref{f:struct1} (c), which has 21 predetermined links. 
The ratio 817190~:~6435 illustrates the point that the number of 
\bfs' with  high $n_{10}$ values is much smaller than the number of
\bfs' with low $n_{10}$ values. 

We now give a heuristic argument showing 
that there is an approximate relation between 
the smallest possible Hamming distance \dhm\ between two 
structures $\bfs_1$ and $\bfs_2$ and the difference in the 
$n_{10}$ values of the two structures, 
\dn=$n_{10}(\bfs_1) - n_{10}(\bfs_2)$; for simplicity we assume 
that $n_{10}(\bfs_1) > n_{10}(\bfs_2)$.   For this discussion we 
ignore the two end points of the structures, so that (on a square 
lattice) all the segments on an \bfs\ partitioned by 0-1 links have 
at least two 0 or two 1 digits. 
We begin by considering the case when $\bfs_2$=$\bfs_1$.  Then both 
\dh\ and \dn\ are zero.  Suppose we can generate $\bfs_2$ by 
swapping the positions of a pair of 0's and a pair of 1's in 
$\bfs_1$ (while keeping in mind that in most cases such 
an operation would not give an \bfs; it would give a \bfp\ 
that is not in \calS).  Then \dh~=~2 and, 
depending on the position of the replaced pair of 0's in $\bfs_1$, 
\dn~=~0 or 2.  Any other pair of $\bfs_2$ and $\bfs_1$ having \dn~=~2 
will have \dh~$>$~2.  Thus \dhm\ is 2 for \dn~=~2. 
Similarly, if we generate $\bfs_2$ by exchanging the 
positions of a pair of 0's and a pair 1's in $\bfs_1$, for example: 
\begin{eqnarray}
     &(&\cdots 0111111110\cdots 1000000001\cdots)\nonumber\\
\to  &(&\cdots 0111111000\cdots 1001100001\cdots)
\label{e:hamming1}
\end{eqnarray}

\begin{eqnarray}
{\rm or}\qquad   &(&\cdots 0111111110\cdots 1000000001\cdots)\nonumber\\
\to  &(&\cdots 0111100110\cdots 1001100001\cdots)
\label{e:hamming2}
\end{eqnarray}
then \dh~=~4 and \dn~=~2 (Eq.(\ref{e:hamming1})) or 4 
(Eq.(\ref{e:hamming2})).  Again any other $\bfs_2$ and $\bfs_1$ 
having \dn~=~2 or 4 will have \dh~$>$~4.  Thus \dhm\ is 4 for \dn~=~4. 
Arguing along this line it can be shown that 
\dhm~$\approx$~\dn.  
\begin{center}
\begin{minipage}{3in}
\begin{figure}[m]
\epsfysize=7.0cm\epsfbox{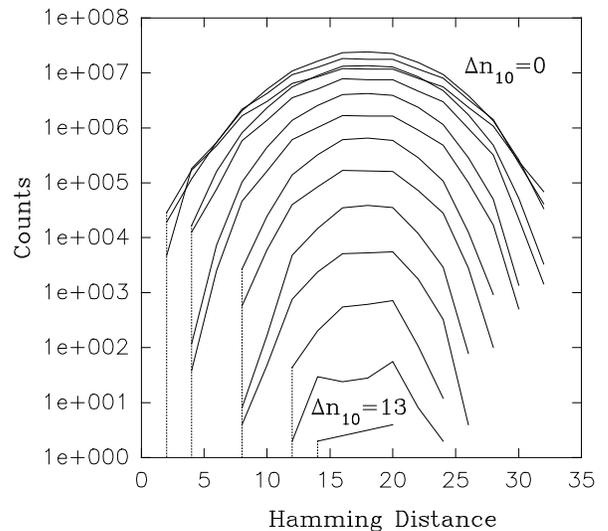}
\caption{The Hamming distances between pairs of all the 30408 structural
sequences on a 6$\times$6 lattice. The vertical dashed lines indicate
the minimal Hamming distances for different $\Delta n_{10}$.}
\label{f:histo}
\end{figure}
\end{minipage}
\end{center}
In Fig.~\ref{f:histo}, the logarithmic distributions of the Hamming 
distances between pairs of \bfs' with fixed values of \dn\ are plotted 
for a 6$\times$6 lattice. 
The relation between \dhm\ and \dn\ is clearly displayed. 
Notice that all distributions peak at a Hamming distance of 15-20, 
with the width of the distribution decreasing monotonically with \dn. 

\begin{center}
\begin{minipage}{3.2in}
\begin{figure}[m]
\epsfysize=9.5cm\epsfbox{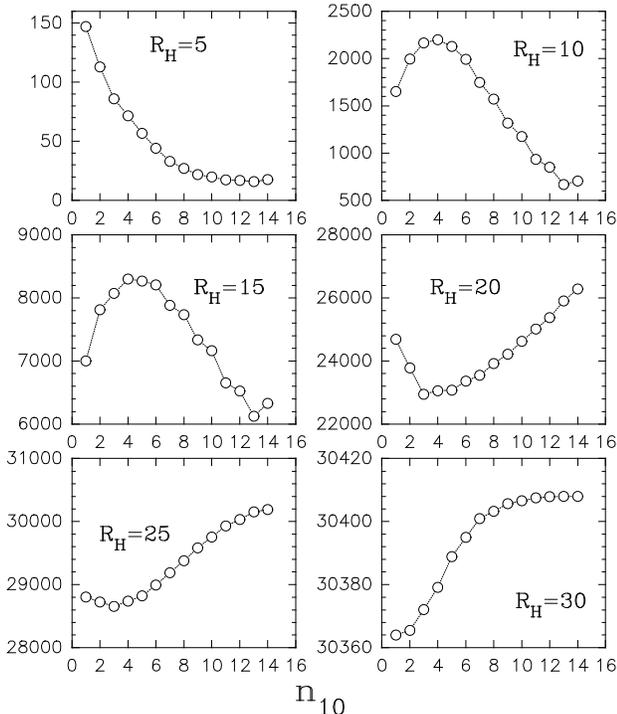}
\caption{Average number of neighboring structures within different
Hamming distances $R_H$ for a 6$\times$6 lattice.}
\label{f:hamm}
\end{figure}
\end{minipage}
\end{center}
\begin{figure}[m]
\epsfysize=6.5cm\epsfbox{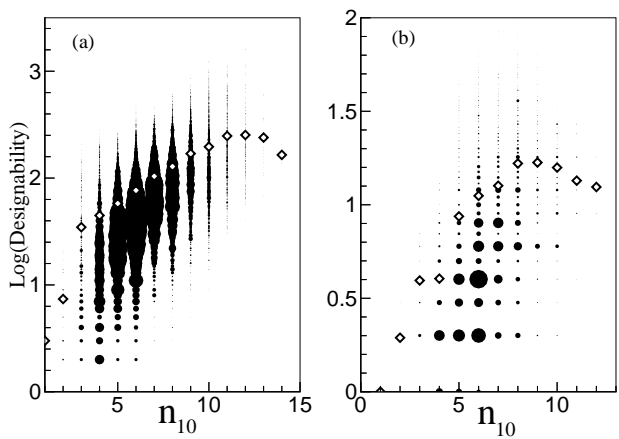}
\caption{Designability distributions for (a) 6$\times$6 square lattice 
and (b) 21-site triangular lattice. See the text for detail.}
\label{f:design}
\end{figure}

It has already been shown that the number of \bfs' with large $n_{10}$
is much smaller than the number of \bfs' with small $n_{10}$. 
Hence the former kinds of \bfs ' will be even more sparsely 
distributed in \calP\ than the latter kinds.   Thus given an  arbitrary 
\bfs\, the chances are that most of its nearest neighbors 
will have relatively small $n_{10}$'s.   An \bfs\ with large 
$n_{10}$ will be farther away from its nearest neighbors 
than if it has a smaller $n_{10}$.   
This is indeed brought out in Fig.~\ref{f:hamm}, where each curve 
plots as a function of $n_{10}$ the number of neighboring
\bfs' in \calS\ within a Hamming distance $R_H$, 
averaged over those \bfs' specified by $n_{10}$.  
It is seen that so long as $R_H \le 15$, 
\bfs' with  large $n_{10}$ has far fewer nearby neighbors (in \calS) 
than \bfs' with smaller $n_{10}$. 
It follows that \bfs' with 
large $n_{10}$ will on average have large Voronoi polytopes, hence 
high designabilities. 
Note that the approximate proportional relation between \dn\ and \dhm\
is not expected to be limited to square lattices although the proportional
constant is expected to be dependent on lattice type. 

\begin{center}
\begin{minipage}{2.7in}
\begin{table}[m]
\caption{$n^{max}_{10}$ and $n^{peak}_{10}$ for several lattices}
\begin{tabular}{ccc} 
lattice & $n^{max}_{10}$ & $n^{peak}_{10}$\phantom{xx}\\
\hline
$4\times 4$ & 6 & 4\phantom{xx}\\
\hline
$4\times 6$ & 9 & 8\phantom{xx}\\
\hline
$5\times 5$ & 10 & 7\phantom{xx}\\
\hline
$4\times 7$ & 11 & 10\phantom{xx}\\
\hline
$5\times 6$ & 12 & 9\phantom{xx}\\
\hline
6$\times$6 & 14 & 12\phantom{xx}\\
\hline
21-site triangle & 12 & 9\phantom{xx}\\
\end{tabular}
\label{t:peak}
\end{table}
\end{minipage}
\end{center}

In Fig.~\ref{f:design} (a) and (b) the logarithmic designability 
is plotted as a function of $n_{10}$ for  a 6$\times$6 square 
lattice and a 21-site triangular lattice, respectively.  
The size of each disc indicates the number of \bfs' having the specific 
$n_{10}$ and designability and an open diamond indicates the 
average designability of all \bfs' having the specified $n_{10}$. 
On the whole the average designability increases with $n_{10}$ up
to near the maximum  $n_{10}$.  For $n_{10}$ near the maximum value 
it appears that the heuristic argument given above breaks down, 
probably partly for boundary effects, and partly because the number 
of structures with the largest values of $n_{10}$ is very small 
(3 for $n_{10}=14$ and 24 for $n_{10}=13$ among the 30408
\bfs~$\in$~\calS\ on a 6$\times$6 square lattice) so that statistical
fluctuations become important.   The designability distributions
on several other lattices were studied and the pattern shown in 
Fig.~\ref{f:design} persisted.   The result is summarized in 
Table~\ref{t:peak}, where $n^{max}_{10}$, the maximum $n_{10}$ and 
$n^{peak}_{10}$, the $n_{10}$ where the largest average designability 
occurs, are given for each lattice.     In all the cases 
$n^{peak}_{10}=n^{max}_{10}-2\pm 1$. 
Results for three-dimensional lattices will be shown elsewhere.

\section{Comparison with Real Proteins}
\label{s:pdb}

It has been shown that the mathematical contents of 
the mean-field HP model and the LS model are essentially identical. 
The physical (or biological) interpretations given to the two 
models are however entirely different.  The mean-field HP model is based 
on the assumption that hydrophobic residues would congregate in the 
core as much as possible.  The LS model is based on the assumption 
that large residues would be excluded from the core as much as possible. 
To see which model is closer to Nature we 
compare the results of the two models with real proteins by matching 
model peptide sequences against protein sequences
culled from data banks.   For either model, the model sequences 
are the two sets of sequences among a total 
$26,000,000$ randomly sampled 36-word binary sequences 
that select the most highly designable and least designable structures,  
respectively, on a 6$\times$6 lattice.  

We consider the frequency distributions of the set of 
sequences \{${\cal P}_\lambda | \lambda = h,l,S,\phi,\alpha,\beta,
\phi',\alpha',\beta'$\}, where the subscript $h$ denotes 
the concatenated 27006 peptides mapped to the 15 
most highly designable structures in the mean-field HP model; 
$l$, the concatenated 24134 peptide sequences mapped to the 1545 least 
designable structures in the mean-field HP model; 
S, the concatenated 22789 peptides mapped to the 364 most 
highly encodable structures in the LS model \cite{lsnote}; 
$\phi$, the concatenated protein sequences in PDB \cite{PDB}, 
converted to a binary sequences based on the hydrophobicity of the 
peptides;
$\alpha$, same as $\phi$, but includes only segments of protein 
sequences that fold to $\alpha$ helices; 
$\beta$, same as $\phi$, but includes only segments of protein 
sequences that fold to $\beta$ sheets; 
$\phi'$, $\alpha'$ and $\beta'$, same as $\phi$, $\alpha$ and 
$\beta$, respectively, except that protein sequences are 
converted to binary ones based on
the volume of residues. 
The ten residues designated polar (P) are: Lys, Arg, His, Glu, 
Asp, Gln, Asn, Ser, Thr, Cys \cite{radzicka88} and 
the ten residues designated as L-type residues 
are, in descending order of volume, 
Trp, Tyr, Phe, Arg, Lys, Leu, Ile, Met, His and Gln  \cite{zamyatin72}. 
That the HP and LS models differ in physical and biological contents 
is predicated by the fact that the two lists overlap poorly. 
This predication will not change if the cut-off points of either or 
both lists are varied slightly. 
The sequences \calP$_h$ and \calP$_S$ will be referred to 
as the most foldable peptides in the HP and LS models, respectively.

To compare the sequences, we employ a Cartesian coordinate
representation for symbolic sequences \cite{hao98}, here applied to 
binary sequences. 
Let $\cal S$ denote the set of $2^l$ binary strings $\sigma$ of 
length $l$.   Given a binary sequence ${\cal P}_\lambda$ 
of length $L$ and a string length $l$ 
(we are interested only in cases when $L>>l$), there is the set 
$\{f^{(l)}_\lambda(\sigma)| \sigma \in \cal S\}$ 
of frequencies of occurrence of the string 
$\sigma$ in $\lambda$.  The frequencies may be obtained, say, 
by counting while sliding a  window $l$ digits wide along $\lambda$. 
The frequency depends on the ratio of 0 to 1 digits in the sequence. 
This ratio, $r_{\lambda}$, is 0.983, 1.039, 0.553, 0.960, 0.993,
0.720, 0.734, 0.917 and 0.934, respectively, for the sequences 
${\cal P}_\lambda$, $\lambda$= $h, l, S, 
\phi,\alpha,\beta, \phi',\alpha',\beta'$. 
In order to make a fair 
comparison of the sequences adjustments need to be made to compensate 
for the disparity in the 0 to 1 ratios.  For this purpose we define 
a normalized frequency $f'$ by 
\begin{equation}
{f'}^{(l)}_\lambda(\sigma) = 
(r_{\lambda})^{n_\sigma} f^{(l)}_\lambda(\sigma)
\label{e:fprime}
\end{equation}
where $n_\sigma$ is the number of 0's in $\sigma$. 
Sequences in the normalized frequency set $\{{f'}^{(l)}_\lambda(\sigma)\}$ 
now have 0 to 1 ratios equal to unity. 
 
In what follows we consider only cases when $l$ is even, $l=2k$. 
Let ${\cal L}$ be a $2^k\times 2^k$ lattice 
with spacing $2^{-k}$, and $\pi$ be a one-to-one mapping 
from $\cal S$ to $\cal L$, $\pi: \cal S \to \cal L$ by: 
\begin{equation}
\pi(\sigma) = (x,y) \equiv \left(\sum^k_{i=1} \sigma_{k+i}\cdot 2^{-i},\ 
\sum^k_{i=1} \sigma_{i}\cdot 2^{-(k-i+1)}\right)
\label{e:metric}
\end{equation}
where $\sigma = [\sigma _1, \sigma _2, 
\cdots, \sigma _{2k}]$ is a string in $\cal S$  and 
$(x,y)$ is a site on  $\cal L$. 
From the set $\{{f'}^{(l)}_\lambda(\sigma)\}$ we define a normalized
relative frequency distribution of $\lambda$ on the lattice $\cal L$: 
\begin{equation}
F^{(l)}_\lambda(x,y)\equiv F^{(l)}_\lambda(\pi(\sigma))
=\left({f'}^{(l)}_\lambda(\sigma)-{\bar f}^{(l)}_\lambda\right)/
Z_\lambda 
\label{e:normalization}
\end{equation}
where ${\bar f}^{(l)}_\lambda$ is the mean frequency and 
\begin{equation}
Z_\lambda= \left(\sum_{\sigma\in \cal S} {f'}^{(l)}_\lambda(\sigma) -
{\bar f}^{(l)}_\lambda\right)^{1/2}
\label{e:norm_rel}
\end{equation}

Figs.~\ref{f:hp3} and \ref{f:ls3} show the distributions 
$F^{(6)}_\lambda$, $\lambda$= $\phi$, $\alpha$, $\beta$ and $h$,  
and $\lambda$= $\phi'$, $\alpha'$, $\beta'$,  and $S$,  respectively. 
In the figures, the magnitude of the distribution is coded into 
the gray scale shown at the top of the figures.  
From the fact that (b) and (d) in Fig.~\ref{f:hp3} have their brightest
and darkest regions, respectively, 
at generally the same locations, it is evident that 
\calP$_h$ ((d)),  the most foldable peptides in the HP-model, is closest 
to \calP$_\alpha$ ((b)), the sequence that represents 
$\alpha$ helix segments in real protein sequences. 
In comparison, although (a) looks similar to (b), it is not so 
similar to (d).  In particular, 
some of the brightest regions in (a) are dark 
in (d), and vice versa.  In sharp contrast (c), which 
represents $\beta$ sheet segments in real protein sequences, is 
entirely different from all the other distributions in Fig.~\ref{f:hp3}.   

Turning to Fig.~\ref{f:ls3}, it is noticed that (d), representing 
the most foldable peptides in the LS model, is very similar to 
its counterpart in the HP model, Fig.~\ref{f:hp3} (d).  This is 
as expected because the mathematical contents of the two models  
are essentially identical.   On the other 
hand, (d) is very dissimilar to (a), which represents all protein 
sequences in PDB, but with the residues partitioned according 
to the LS model.  This shows that size of the residue is not the most 
dominant factor in protein structure. 

The frequency distributions shown in Figs.~\ref{f:hp3} and \ref{f:ls3}  
are repeated in Figs.~\ref{f:hp4} and \ref{f:ls4}, except that 
the word length $l$ is now eight instead of six.  This implies that  
the sequences $\cal P_{\lambda}$ are now examined with a finer 
resolution.  The result is similar to the $l=6$ case: 
the most foldable peptides in the HP model closely resemble the 
$\alpha$ helix segments of real protein, while the foldable peptides 
in the LS model do not resemble real proteins. 

\begin{center}
\begin{minipage}[m]{2.7in}
\begin{figure}[m]
\epsfysize=6.5cm\epsfxsize=7.5cm\epsfbox{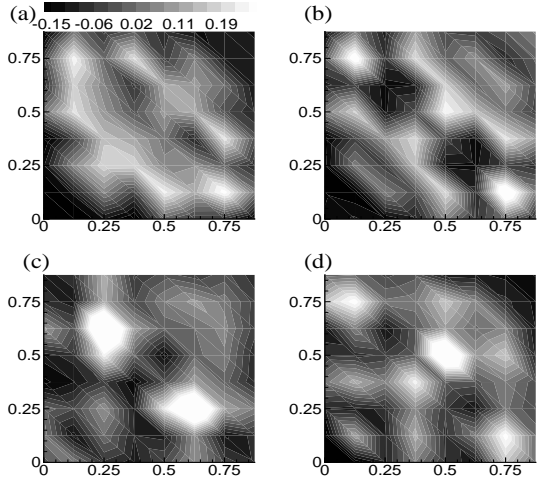}
\caption{Frequency distributions  of strings of length 6 in the 
sequences (a) \calP$_\phi$, (b) \calP$_\alpha$,
(c) \calP$_\beta$, and (d) \calP$_h$; see text for description.}
\label{f:hp3}
\end{figure}
\vspace{-10pt}
\begin{figure}[m]
\epsfysize=6.5cm\epsfxsize=7.5cm\epsfbox{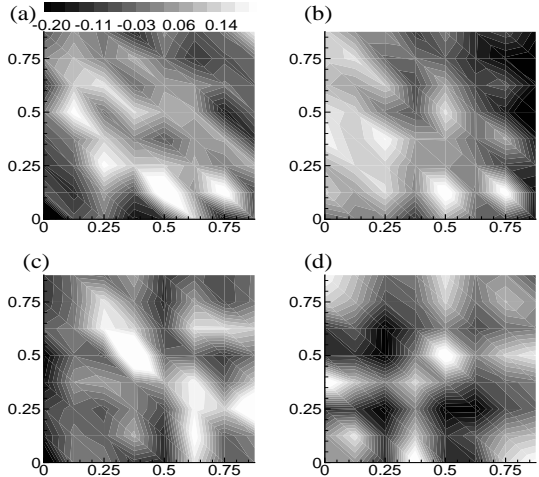}
\caption{Frequency distributions  of strings of length 6 in the 
sequences (a) \calP$_\phi'$, (b) \calP$_\alpha'$,
(c) \calP$_\beta'$, and  (d) \calP$_S$; see text for description.}
\label{f:ls3}
\end{figure}
\end{minipage}
\end{center}
\medskip

The sequences $\cal P_{\lambda}$ may be compared in a more 
quantitative manner through the overlap of frequency distributions:
\begin{equation}
O^{(l)}_{\lambda\lambda'}= \sum_{\sigma\in \cal S} 
F^{(l)}_\lambda(\pi(\sigma))F^{(l)}_{\lambda'}(\pi(\sigma)).
\label{e:overlap}
\end{equation}
The overlaps $O^{(l)}_{\lambda\lambda'}$, for a number 
of pairs ($\lambda,\lambda'$) selected from the set 
\{$h,l,S,\phi,\alpha,\beta, \phi',\alpha',\beta'$\}, and  for
$l= 4\sim 14$ are given in Fig.~\ref{f:overlap}. 

One first notices that, with the exception of $O^{(l)}_{h S}$ 
(\Bull~ in Fig.~\ref{f:overlap}), all the 
overlaps approach zero as the word length $l$ increases.  This 
is so because the resolving power of the method  increases 
with $l$; for sufficiently large $l$, the resolution becomes so 
large that any two sequence that does not have substantial 
and extended sequence identity will have zero overlap. 
That $O^{(l)}_{h S}$ has large positive correlation throughout 
the whole range of $l$ studied is expected from the mathematical 
equivalence of the HP and LS models.  
In Ref. \cite{micheletti98}, the parameter $a$ in Eq.(\ref{e:lsmodel}) 
was taken to be infinity to emphasize the steric 
constraint on the residues.   Here we had done the same just to 
conform to Ref. \cite{micheletti98}. 
On the other hand, since in the present study all the structures 
are self-avoiding paths on a discrete lattice, the steric constraint
caused by the existence of the backbone is automatically satisfied.  
Therefore, so far as the intention of the LS model is concerned, 
a small and positive, but not infinite, value for $a$ would have 
sufficed.  

\begin{center}
\begin{minipage}[m]{3in}
\begin{figure}
\epsfysize=6.5cm\epsfxsize=7.5cm\epsfbox{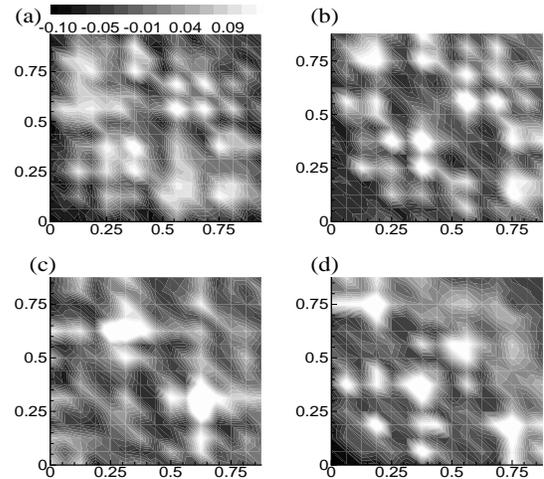}
\caption{Frequency distributions  of strings of length 8 in the 
sequences (a) \calP$_\phi$, (b) \calP$_\alpha$,
(c) \calP$_\beta$, and (d) \calP$_h$.}
\label{f:hp4}
\end{figure}
\vspace{-10pt}
\begin{figure}[m]
\epsfysize=6.5cm\epsfxsize=7.5cm\epsfbox{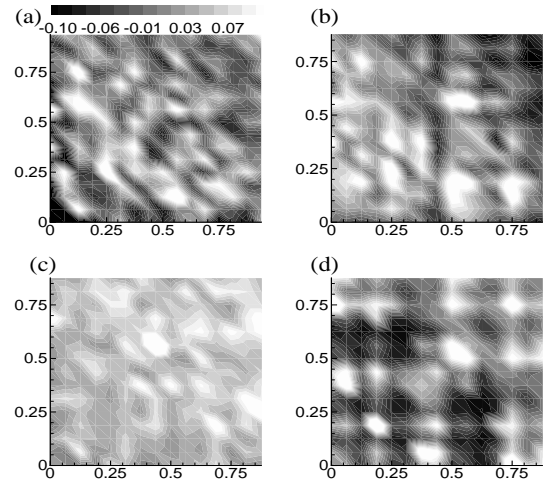}
\caption{Frequency distributions  of strings of length 8 in the 
sequences (a) \calP$_\phi'$, (b) \calP$_\alpha'$,
(c) \calP$_\beta'$, and (d) \calP$_S$.}
\label{f:ls4}
\end{figure}
\end{minipage}
\end{center}
\medskip

The overlap $O^{(l)}_{\phi\alpha}$ (filled $\bigtriangleup$) 
is larger than most other overlaps for 
much of $l$'s shown in the figure.  This is connected to a basic fact 
of proteins: $\alpha$ helices account for almost half of the total amount 
of protein sequences in PDB.  The overlap drops sharply when $l$$\ge$12 
because most $\alpha$ helix segments are shorter than 15 residues 
long. 

Next in order of magnitude are the overlaps $O^{(l)}_{\alpha h}$ and 
$O^{(l)}_{\phi h}$ (filled $\bigtriangledown$ and {\Large$\bullet$}); 
these have large positive values for the smaller 
$l$'s.   This reveals that the mean-field HP model 
provides a coarse-grained description of some features of the real 
proteins and suggests that the basic assumption 
of the model - that local residue-water interaction is the dominant 
cause for protein folding - is consistent with the mechanism for 
the formation of $\alpha$ helices.   The overlaps decrease with 
increasing $l$ for the general reason given above.  
On the other hand, the negative correlation shown by the 
negative value of the overlap $O^{(l)}_{\beta h}$  
($\bigtriangledown$)  shows that the 
same assumption is inconsistent with what causes the formation 
of $\beta$ sheets.   Two of the obvious reasons are: 
whereas most $\beta$ sheets are buried in the interior of proteins, 
the mean-field HP model differentiates only surface from core sites 
but has no means of influencing the interior structure of proteins; 
the stability of most $\beta$ sheets depends on long-range interactions 
that are absent in the model. 
 
\begin{center}
\begin{minipage}[m]{3in}
\begin{figure}[m]
\epsfysize=6.5cm\epsfxsize=7.5cm\epsfbox{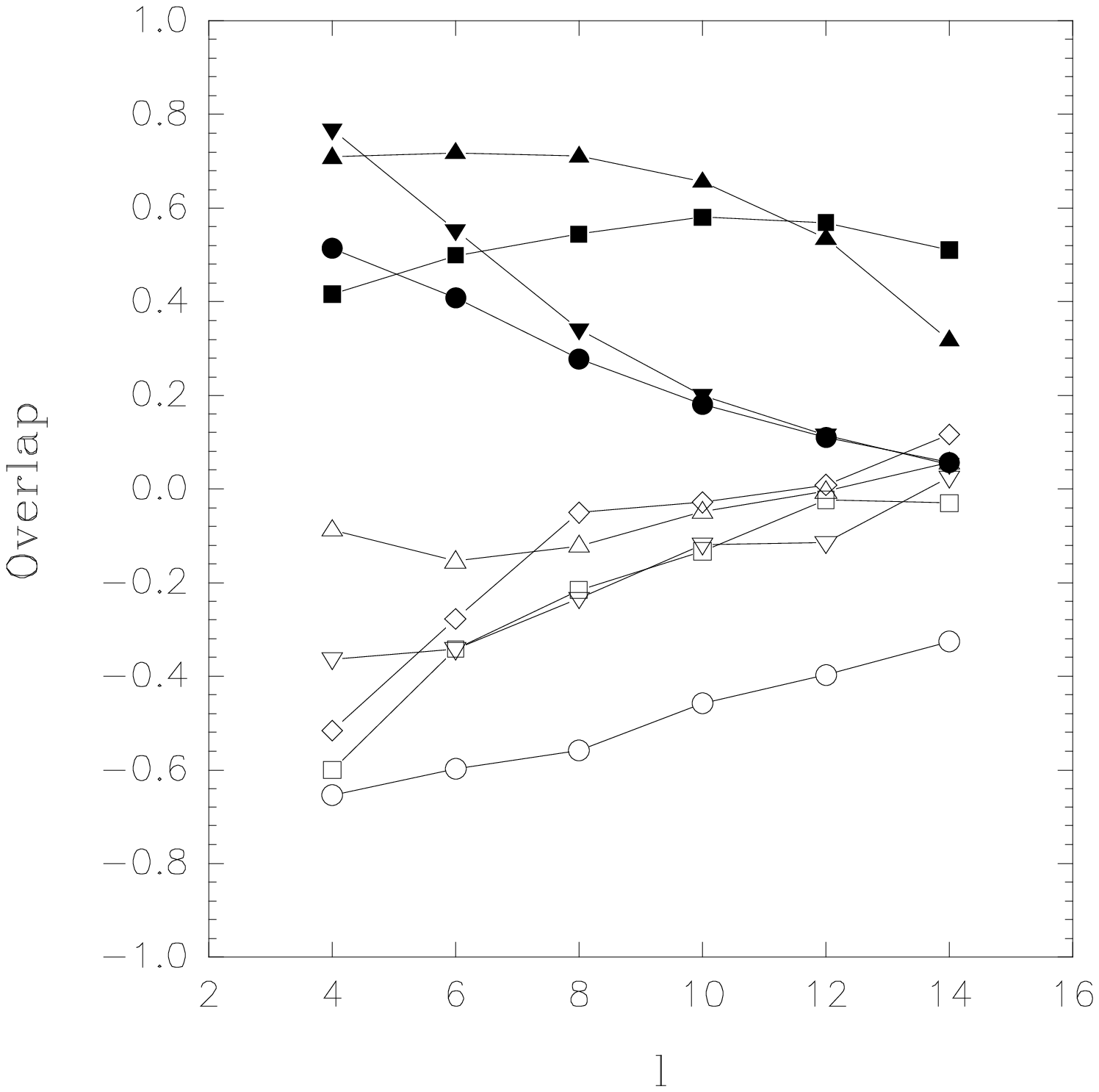} 
\caption{Overlap of frequency distribution functions 
versus word length $l$:
$O^{(l)}_{\phi\alpha}$ (filled $\bigtriangleup$), 
$O^{(l)}_{\alpha h}$ (filled $\bigtriangledown$), 
$O^{(l)}_{\phi h}$ ({\Large$\bullet$}),
$O^{(l)}_{h S}$ (\Bull), 
$O^{(l)}_{\alpha' S}$ ($\bigtriangleup$), 
$O^{(l)}_{\beta h}$ ($\bigtriangledown$),
$O^{(l)}_{\beta' S}$ ({\large$\Diamond$}),
$O^{(l)}_{\phi' S}$ ({\large$\Box$})  
and $O^{(l)}_{h l}$ ({\small$\bigcirc$}).
See text for the description of the subscripts 
$h, l, S, \phi, \alpha, \beta, \phi', \alpha'$ and $\beta'$.}
\label{f:overlap}
\end{figure}
\end{minipage}
\end{center}

The negative value of the overlaps between 
\calP$_S$ and \calP$_{\phi',\alpha',\beta'}$ 
({\large$\Box$}, $\bigtriangleup$ and {\large$\Diamond$}, respectively) 
indicates that 
the highly foldable peptide sequences in the LS model are 
anti-correlated with the real protein sequences for $l\le 6$ and
uncorrelated for larger $l$.  This confirms what is already seen in 
Figs.~\ref{f:ls3} and \ref{f:ls4}: 
that size effect is not the dominant factor determining the 
formation of a stable protein conformation.  
Finally, the large negative values of the overlap $O^{(l)}_{h l}$ 
({\small$\bigcirc$}) for all values of $l$ tested simply verify that 
the most and least foldable peptides in the HP model are highly 
dissimilar however they are compared. 

\section{Discussion}
\label{s:discussion}

Because conformation designability in protein structure refers to the
natural selection of a very small number of topological classes of
native conformations over the vast total number of classes, it is a
topic that can be suitably studies in coarse-grained settings
such as in lattice models.  Previous lattice model studies have firmly
established that indeed only a very small number of (model)
structures, out of a very large total number, are highly designable.
It has not been shown why this phenomenon should arise, and to what
classes of native conformations would the highly designable structures
correspond.  In this paper, taking advantage of the geometric picture
for the designability problem given in \cite{li98}, namely that 
designability of a structure in the mean-field HP 
model is proportional to Voronoi volume of
that structure in a certain hyperspace, we showed that uneven
designability arises because a type of structures - those with the
largest numbers of surface-core switchbacks - are very rare, and that
their nearest neighbors in the hyperspace are other similar rare
structures.  Hence such structures have the largest Voronoi volumes and
the highest designabilities.  Because the hyperspace of structures has
properties independent of the two-dimensional lattices used in the
present study, this conclusion is expected to stand for other more
realistic lattices.  Indeed, the same effect was observed on a
three-dimensional lattice based on an icosahedron \cite{wangbx}.

The identification of structures having the largest numbers of
surface-core switchbacks with the conformation classes of observed
proteins entails certain physical and biological implications.
Proteins choosing such structures as native conformations would tend
to have ratios of numbers of H-type and P-type residues close to being
unity.  Indeed, the averages of H to P ratios for all the protein
sequences in PDB, for the segments that folds to $\alpha$ helices and
and for those that fold to $\beta$ sheets, respectively, are all very
close to unity.  Proteins having structures with many surface-core
switchbacks are expected to be energetically favored.  For such
proteins would by and large have alternating P and H residues that
match the pattern of the structures, and the outward-pointing force
exerting on the P-type residues and the inward-pointing force exerting
on the H-type residues together would make the protein especially
sturdy.

On the mean-field HP lattice, high-designability structures tend not
to have long sequences of contiguous sites that are purely core sites
or purely surface sites (see Table~\ref{t:string} in Appendix), because such
structures tend to be involved in degenerate cases - peptides with
corresponding contiguous subsequences of P- or H-type residues 
(or S- or L-type residues in the LS model) would
easily have two or more such structures as ground states - and for
that reason the peptide and the degenerate structures 
would have been excluded from the set of allowed peptides and acceptable 
structures, respectively.  
This practice is justified biologically: peptides and conformations 
involved in degeneracy (in a coarse-grained sense)  are presumably 
filtered out by evolution because they would make for functionally 
unreliable proteins.  In fact, relatively few
proteins in PDB have sequences containing long segments of
contiguous P- or H-type residues whose native conformations have 
long segments of contiguous surface or buried sites \cite{PDBstats}.  Such
native conformations are presumably generated by the finer details of
inter-residual interactions, and the conformation classes to which they
belong would not have counterparts among the high designability
structures given by simple, coarse-grained lattice models.

Because structures on square lattices are not realistic enough for
direct comparison with empirically observed topological conformation
classes, we compared model peptides folding into such structures,
namely the most foldable peptides, with (binarized) peptide sequences
in the PDB.  If the highly designable structures are rich in 
surface-core switchbacks then the highly foldable peptides should 
be rich in H and P singlets and HH and PP doublets.   
In Table~\ref{t:string} in the Appendix it is seen that the the highly
foldable peptides in the mean-field HP model are rich in HHPP (or PPHH) 
but poor in HP (or PH) repeats.  
This reflects an artifact of the square lattice.  On
such lattices, the shortest surface-core switchback motif is
surface-surface-core-core (or core-core-surface-surface) repeats 
while surface-core repeats do not exist (see first two ``constraints'' 
in Section \ref{s:geometric}).  We showed that the most foldable
peptides match well with those segments of protein sequences in PDB
that fold into $\alpha$ helices but match relatively poorly with
segments that fold into $\beta$ sheets.  $\alpha$ helices are most
commonly amphipathic and lie on the outside of their host proteins.
With 3.6 residues per turn, such $\alpha$ helices tend to change from
H to P residues with a periodicity of three to four.  That is, they
should have a predominance of alternating HH and PP doublets
interspersed with H and P singlets.  Indeed, of all peptide sequences
that code $\alpha$ helices in the PDB, 24\% of H to P (or P to H)
changes are after singlets, 36\% are after doublets and 22\% are after
triplets.  This implies that $\alpha$ helices are relatively rich in
HHPP repeats and this could explain why the most foldable model
peptides (in the mean-field HP model) match well with $\alpha$
helices.

The situation is different with respect to $\beta$ sheets.  The most
common domain structures in proteins are $\alpha$/$\beta$ domains that
consist of a central group of $\beta$ sheets surrounded by $\alpha$
helices.  The $\beta$ sheets in these domains will not be rich in
either HHPP or HP repeats.  In the second large group of protein
domain structures, comprised of antiparallel $\beta$ sheets, some of
the sheets are on the outside of the protein and these are rich in HP
repeats but not in HHPP repeats.  A superfamily of proteins containing
such $\beta$ sheets has members such as the human plasma
retinal-binding protein and $\beta$-lactoglobulin, a protein that is
abundant in milk.  Of all peptide sequences that code $\beta$ sheets
in the PDB, 33\% of H to P (or P to H) changes are after singlets,
28\% are after doublets and 18\% are after triplets.  Hence the most
foldable model peptides would match poorly with $\beta$ sheets.

If our computation were carried out on a lattice that allowed
structures with surface-core repeats then the foldable model peptides
would have better matched sequences coding for $\beta$ sheets.  
Still, because the only
interaction taken into account in the mean-field HP model is the
hydrophobicity of the residues, whereas the formation of the majority
of $\beta$ sheets depend on other details of interesidual
interactions, we cannot expect the most foldable model peptides to
have a good match with the majority of $\beta$ sheets irrespective
of what lattice was used.

If  hydrophobicity but not inter-residual interaction 
is indeed the main force that drives the 
formation of $\alpha$ helices, then we can better understand why 
$\alpha$ helices are formed on a time scale 
of the order $10^{-7}$s \cite{Munoz,Williams}, right after the 
collapse of the protein to globular shape, and why  it 
takes ten times longer for the formation of $\beta$ sheets,
which involves interactions between residues distantly separated 
on the primary structure.   
  This scenario is 
consistent with the finding in a recent statistical analysis of
experimental data: local contacts play the key role in
fast processes during folding \cite{plax98}.

We have shown that the mathematical content of the LS model, which
partitions residues into large (L) and small (S) ones, was essentially
the same as that of the mean-field HP model.  Hence the binary
composition of the most foldable peptides in the two models are quite
similar (see Table~\ref{t:string}, Appendix).  However, because not all large
(small) residues are hydrophilic (hydrophobic), the most foldable
peptides in the two models are mapped to significantly different sets
of (binarized) protein sequences.  The result is that the most
foldable peptides in the LS model do not match well with any subset of
proteins in the PDB.  This means that steric hindrance effect arising
from different sizes of the residues is not the main driving force for
protein folding.

We thank the National Center for High-Performance Computing 
(NCHC) for providing support in computation and accesses to PDB.
This work is partly supported by grants NSC89-2213-E-321-004 
to ZYS, NSC89-M-2112-008-0022 to HCL and NSC87-M-2112-007-004 to BLH
from the National Science Council.   HCL thanks the Physics Department 
of Stanford University where this work was partly written.


\bigskip
\begin{center}
{\bf APPENDIX}
\end{center}

Here we show how the two lattice models differ by 
comparing strings of several lengths that have the highest and 
lowest frequencies of occurrence, called the most and least favored 
strings, respectively, in the sequences \calP$_h$ 
and \calP$_S$, which are the concatenated sequences of the 
mostly highly foldable peptides in the 
mean-field HP and LS models, respectively.  
In Table~\ref{t:string}, the first and sixth columns list such 
strings.  Strings of different lengths are ranked separately 
by their normalized relative frequency of occurrence 
(Eq.~(\ref{e:norm_rel})); the string with the highest (lowest) 
frequency is ranked 1 ($2^l$).  
By definition, an unfavored string has negative frequency. 
Table~\ref{t:string} shows that the most favored strings are 
quite well correlated in the two models but the least favored 
strings are not so. 
It is seen that among 4-mers the repeats (0011) are the most favored 
pattern in both models, long repeats of 1's and 0's are the 
least favored string patterns in the HP model 
favored string patterns
in the HP model and (01) is the the least favored string 
repeat in the LS model.
The reason for this is clear: 
(0011) repeats are the favored pattern in most highly designable structures 
in both models and each of the (peptide) strings (0000), (1111) and (0101) 
is separated from (0011) by the greatest {\it frame independent} 
Hamming distance.
There is an additional disincentive for a peptide to have
(01) repeats in the LS model.   On a square 
lattice such repeats do not appear in a structure sequence, hence, 
with L-type residues (represented by 0 digits) 
strictly forbidden on core sites (represented by 1 digits), a peptide 
string with 01 repeats can only occupy a structure sequence 
composed entirely of surface sites.  This gives the peptide zero binding
energy in the LS model.  The situation is different in the HP model. 
There a peptide string with 01 repeats can occupy a structure sequence 
with 0011 repeats and non-zero binding energy. 

\vspace{0.5in}
\begin{minipage}{6.4in}
\begin{table}
\caption{Strings most and least favored in the mean-field HP and
LS models. Strings of different lengths are ranked separately; 
e.g., the least favored string of length 4 is ranked $2^4$=16.}
\begin{tabular}{|c|cccc||c|cccc|}
Strings most/least & 
\multicolumn{2}{c}{\underbar{\phantom{xx}HP model\phantom{xx}}}&
\multicolumn{2}{c||}{\underbar{\phantom{xx}LS model\phantom{xx}}}&
Strings most/least & 
\multicolumn{2}{c}{\underbar{\phantom{xx}LS model\phantom{xx}}}&
\multicolumn{2}{c|}{\underbar{\phantom{xx}HP model\phantom{xx}}}\\
favored in HP model&freq. &rank&freq. &rank&
favored in LS model&freq. &rank&freq. &rank\\
\hline\hline
 (0110) &  0.4459  & 1 & -0.0468 & 10 & (0011) & 0.3834  & 1 & 0.4272 & 2 \\
\hline
 (0011) & 0.4272  & 2  & 0.3834 & 1 & (1100) & 0.3693 & 2 & 0.4224 & 3 \\
\hline
 (0000) & -0.3883 & 15 & 0.2732 & 3 & (1010) & -0.3815 & 15 & -0.1572 & 11 \\
\hline
 (1111) & -0.3903 & 16 & 0.0109 & 9 & (0101) & -0.3892 & 16 & -0.1594 & 12 \\
\hline\hline
 (001100) & 0.4605 & 1 & 0.2694 & 1 & (001100) & 0.2694  & 1 & 0.4605 & 1 \\
\hline 
 (011001) & 0.2746 & 2 & 0.0656 & 20 & (000011) & 0.2694 & 2 & 0.0515 & 18\\
\hline
 (100110) & 0.2698 & 3 & 0.0672 & 19 & (110000) & 0.2680 & 3 & 0.0369 & 23\\
\hline
 (000001) & -0.1725 & 62 & 0.0379 & 22 & (101010) & -0.2186 & 62 & -0.1253 & 58 \\
\hline
 (100000) & -0.1741 & 63 & 0.0385 & 21 & (010101) & -0.2222 & 63 & -0.1234 & 57 \\
\hline
 (000000) & -0.2694 & 64 & 0.0274 & 25 & (001010) & -0.2224 & 64 & -0.0589 & 39 \\
\hline\hline
 (00110011) & 0.2101 & 1 & 0.1016 & 19 & (11000011) & 0.2318 & 1 & 0.1875 & 4\\
\hline
 (01100110) & 0.2089 & 2 & 0.0541 & 51 & (00001100) & 0.2141 & 2 & 0.1332 & 15\\
\hline
 (11001100) & 0.1977 & 3 & 0.1001 & 20 & (00110000) & 0.2110 & 3 & 0.1191 & 23\\
\hline
 (11000011) & 0.1875 & 4 & 0.2318 & 1 & (00111100) & 0.1684 & 4 & -0.0466 & 200\\
\hline
 (00000011) & -0.0927 & 253 & 0.0293 & 74 & (01010100) & -0.0989 & 253 & -0.0401 & 180 \\
\hline
 (00000001) & -0.1015 & 254 & 0.0301 & 72 & (01010010) & -0.1008 & 254 & -0.0418 & 188 \\
\hline
 (10000000) & -0.1023 & 255 & 0.0334 & 63 & (01001010) & -0.1013 & 255 & -0.0436 & 194 \\
\hline
 (00000000) & -0.1060 & 256 & 0.0088 & 94 & (00101010) & -0.1017 & 256 & -0.0379 & 172 \\
\hline\hline
 (0011001100) & 0.1682 & 1 & 0.902 & 14 & (0011000011) & 0.1837 & 1 & 0.1400 & 4 \\
\hline
 (1100001100) & 0.1574 & 2 & 0.1830 & 2 & (1100001100) & 0.1830 & 2 & 0.1574 & 2 \\
\hline
 (0110000110) & 0.1548 & 3 & 0.1335 & 3 & (0110000110) & 0.1335 & 3 & 0.1548 & 3 \\
\hline
 (0011000011) & 0.1400 & 4 & 0.1837 & 1 & (1001100001) & 0.1230 & 4 & 0.1211 & 8 \\
\hline
 (1111000000) & -0.0408 & 1021 & 0.0220 & 214 & (0101001010) & -0.0441 & 1021 & -0.0173 & 693\\
\hline
 (1110000000) & -0.0414 & 1022 & 0.0508 & 58 & (0100001010) & -0.440 & 1022 & -0.0102 & 528\\
\hline
 (0000000000) & -0.0426 & 1023 & -0.0219 & 773 & (0101010101) & -0.0444 & 1023 & 0.0268 & 893\\
\hline
 (1111111111) & -0.0427 & 1024 & -0.0358 & 914 & (1010101010) & -0.0446 & 1024 & 0.0250 & 869\\
\end{tabular}
\label{t:string}
\end{table}
\end{minipage}


\begin{references}
\bibitem{PDBstats} In the peptide obtained from concatenating all 
the protein sequences in PDB, 
the average length of contiguous same-type residues is
approximately 1.8 residues, with a standard deviation of 1.1
residues.  The total number of residues involved in same-type contigs 
longer than four residues is about 9\% of the total number of residues.
\end{references}
\end{document}